\documentclass[prl, twocolumn, superscriptaddress, amsmath,amssymb]{revtex4-2}

\usepackage{graphicx}
\usepackage{dcolumn}
\usepackage{bm}

\begin{document}

\title{Quantum spin Hall phase in GeSn heterostructures on silicon}

\author{B. M. Ferrari}
\affiliation{Dipartimento di Scienza dei Materiali, Universit\`a di Milano-Bicocca, LNESS and BiQuTe, via Cozzi 55, 20125 Milano, Italy}

\author{F. Marcantonio}
\affiliation{Dipartimento di Scienza dei Materiali, Universit\`a di Milano-Bicocca, LNESS and BiQuTe, via Cozzi 55, 20125 Milano, Italy}

\author{F. Murphy-Armando}
\affiliation{Tyndall National Institute, Cork T12R5CP, Ireland}

\author{M. Virgilio}
\affiliation{Dipartimento di Fisica, Universit\`a di Pisa, Largo Pontecorvo 3, I-56127 Pisa, Italy}

\author{F. Pezzoli}
\altaffiliation{fabio.pezzoli@unimib.it}
\affiliation{Dipartimento di Scienza dei Materiali, Universit\`a di Milano-Bicocca, LNESS and BiQuTe, via Cozzi 55, 20125 Milano, Italy}

\begin{abstract}
Quantum phases of solid-state electron systems look poised to sustain exotic phenomena and a very rich spin physics. We propose a practical silicon-based architecture that spontaneously sustains topological properties, while being fully compatible with the high-volume manufacturing capabilities of modern microelectronic foundries. Here we show how $\mathrm{Ge_{1-x}Sn_x}$ alloys, an emerging group IV semiconductor, can be engineered into junctions that demonstrate a broken gap alignment. We predict such basic building block undergo a quantum phase transition that can elegantly accommodate the existence of gate-controlled chiral edge states directly on Si. This will enable tantalizing prospects for designing integrated circuits hosting quantum spin hall insulators and advanced topological functionalities.
\end{abstract}


\maketitle

The hybridization between valence and conduction states at surfaces and interfaces has been recognized as a key-enabling solution for the realization of topological protected states \cite{bernevig06, konig07,liu08, hasan10}. In heterostructures possessing a broken-gap (BG) alignment of the band edges, electron and hole subbands can be localized at opposite sides of the junctions, while simultaneously retaining the inverted ordering regime \cite{liu08}. This leads to the emergence of helical boundary states, which are characteristic fingerprints of the so-called quantum spin Hall (QSH) phase \cite{bernevig06, konig07, knez11, du15, miao12, zhang13, xu13, lodge21}. Challenges, however, remain before these concepts can be harnessed for their practical use in everyday applications. Were QSH capabilities realised in group IV materials, this would bring  exceptional topological properties into mass-produced devices. Yet, the crucial BG band line-up is the missing ingredient that presently hampers the achievement of genuine quantum phases in such systems.

In this Letter we introduce group IV architectures that will enable quantum phase transitions in ubiquitous Si-based devices. More precisely, we discover the coveted BG in heterojunctions based on  $\mathrm{Ge_{1-x}Sn_x}$ and single out, by the model-solid theory \cite{van89}, unique designs capable of sustaining the QSH phase. We leverage multiband $k \cdot p$ calculations to fully capture the topological phase transitions in specific $\mathrm{Ge_{1-x}Sn_x/Ge_{1-y}Sn_y}$ superlattices (SL). We show that strain and quantum confinement offer practical degrees of freedom to engineer the band inversion. Finally, we extend our numerical investigation to finite size SLs connected to gate leads using self-consistent modelling. In so doing, we explore realistic device geometries and validate the advantageous addition of the electrical tunability owned by these unique topological architectures. 

Lately, alloying Ge with Sn has increasingly attracted attention because it provides a means to carefully induce a transition of the fundamental energy gap \cite{soref93, chang10, wirths16}. Present-day epitaxy of $\mathrm{Ge_{1-x}Sn_x}$ on Si is well established and industrial techniques have already succeeded in synthesizing materials with Sn contents above 30 at.\% \cite{xu19, zheng18}: well beyond the equilibrium solid solubility of the constituent elements, i.e., less than 1\%.
Here we envision an heteroepitaxial stack grown on the so-called virtual substrate (VS), which accommodates the mismatch between the lattice parameters ($\mathrm{a_{Sn}>a_{Ge}>a_{Si}}$), and consists of a $\mathrm{Ge_{1-s}Sn_s}$ buffer deposited on a common Si(001) wafer. The topmost VS film is a relaxed alloy, which can then be used as a template for the repeated deposition of $\mathrm{Ge_{1-x}Sn_x/Ge_{1-y}Sn_y}$ bilayers in a state of biaxial tension or compression, owning to the lattice mismatch with the selected VS. The most feasible method to fabricate VSs would be a standard compositionally graded film \cite{dou18, dou18b, liu21}, but other strategies have been also shown viable \cite{assali19}.

\begin{figure}
   \centering
   \includegraphics[width=8.6cm]{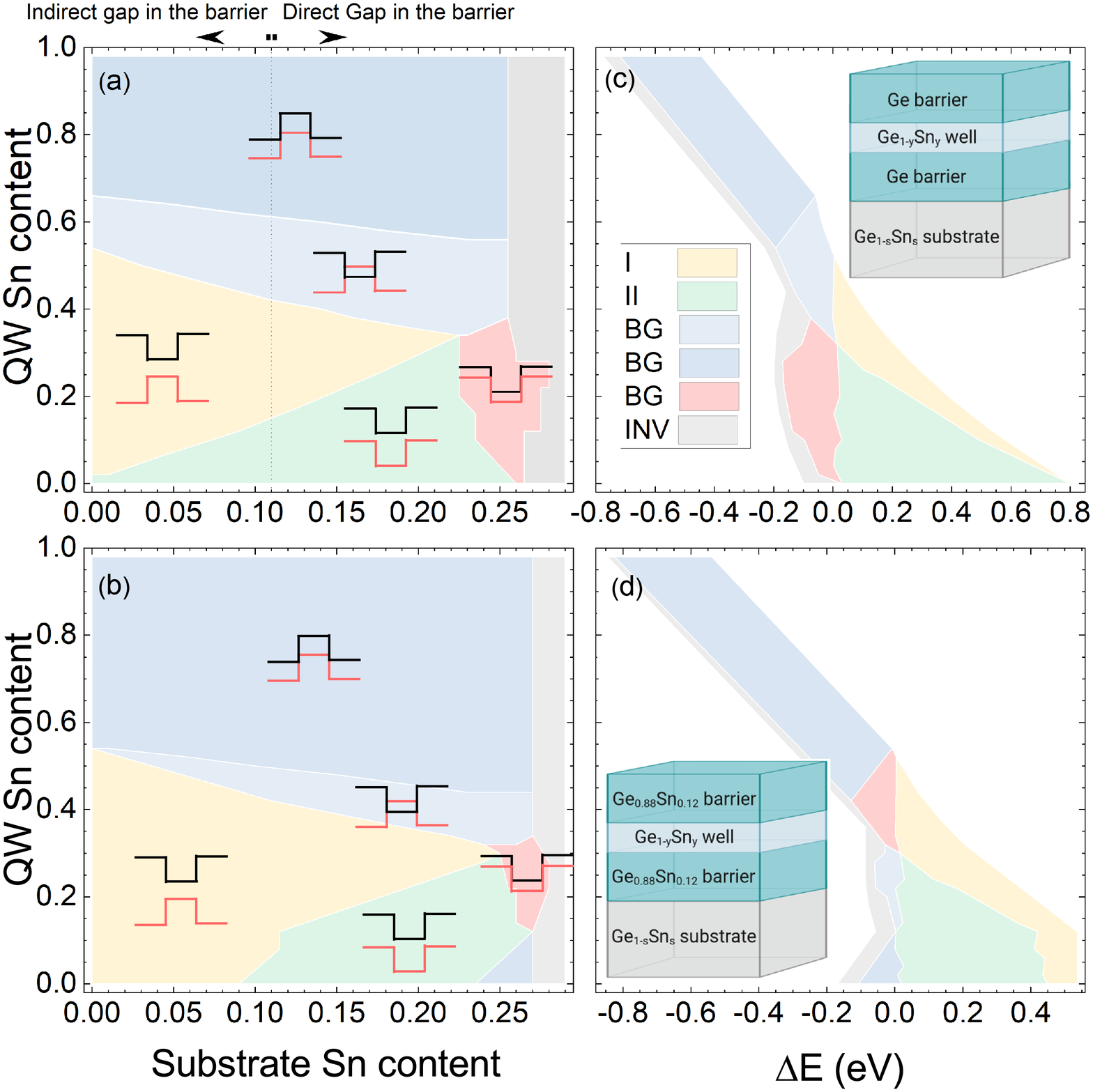}
   \caption{Band-edge profiles of Ge/$\mathrm{Ge_{1-y}Sn_y}$/Ge (a) and $\mathrm{Ge_{0.88}Sn_{0.12}/Ge_{1-y}Sn_{y}/Ge_{0.88}Sn_{0.12}}$ (b) heterojunctions as a function of the Sn concentration of the relaxed substrate. The well/barrier stack is coherent with respect to the lattice spacing pertaining to the $\mathrm{Ge_{1-s}Sn_s}$ substrate. The schematics demonstrate a type I (yellow), type II (green), and broken gap (BG red, light and dark blue) alignments. The region where the band inversion occurs within the same layer as in bulk topological materials is also shown (INV, grey). In (b) the barriers feature a direct gap regardless of the Sn content of the substrate. Minimum energy difference ($\Delta E$) between conduction and valence bands across Ge/$\mathrm{Ge_{1-y}Sn_y}$/Ge (c) and $\mathrm{Ge_{0.88}Sn_{0.12}/Ge_{1-y}Sn_y/Ge_{0.88}Sn_{0.12}}$ (d). Schematics of the heterostructures (not to scale) are shown as insets.} \label{fig1}
\end{figure}

\begin{figure}
   \centering
   \includegraphics[width=8.6cm]{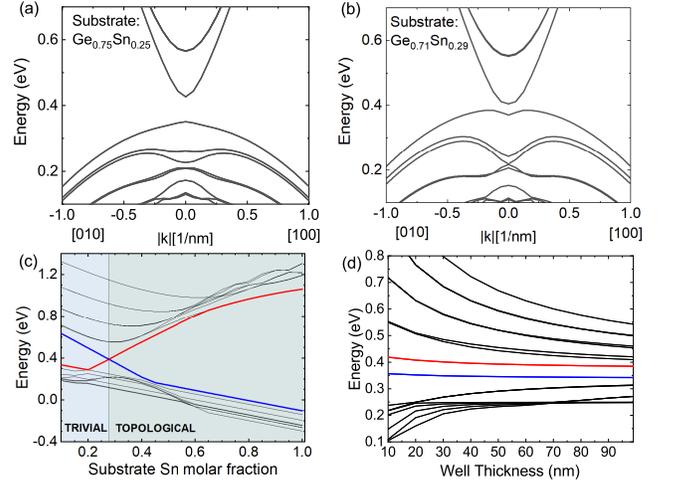}
   \caption{Band structure of a $\mathrm{Ge_{0.78}Sn_{0.22}}$(10 nm)/$\mathrm{Ge_{0.88}Sn_{0.12}}$(25 nm) superlattice (SL) coherently grown on relaxed buffer layers, namely $\mathrm{Ge_{0.75}Sn_{0.25}}$ (a) and $\mathrm{Ge_{0.71}Sn_{0.29}}$ (b). (c) Energy levels of a 10 nm thick $\mathrm{Ge_{0.78}Sn_{0.22}}$ QW embedded in $\mathrm{Ge_{0.88}Sn_{0.12}}$ barriers of 25 nm as a function of the Sn molar fraction of the $\mathrm{Ge_{1-s}Sn_s}$ template. The lowest subband energies for the conduction and valence bands are indicated as red and blue lines. (d) Energy of the confined states of a $\mathrm{Ge_{0.78}Sn_{0.22}}$ QW embedded in $\mathrm{Ge_{0.88}Sn_{0.12}}$ barriers as a function of the well thickness. The substrate is a relaxed $\mathrm{Ge_{0.7}Sn_{0.3}}$.}
   \label{fig2}
\end{figure}

To disclose a rational design of QSH insulators on Si we begin by considering the band alignment established at the $\mathrm{Ge_{1-x}Sn_x/Ge_{1-y}Sn_y}$ heterointerface. We assume the films to be in full registry (coherent) with the lattice of a fully relaxed $\mathrm{Ge_{1-s}Sn_s}$ buffer. To minimize the computational burden, we rely on the flexible approach offered by the well-established model-solid theory that enable us to compute interface line-ups by including strain effects \cite{van89}. The parameters and validation of the model are provided in the Supplemental Material \cite{supp}. Another important point is that we will focus on strained films that are predicted to have a direct band gap. This can provide a viable solution to avoid the competitive presence of absolute minima at the zone edges, like the characteristic L-valleys in relaxed Ge, that could impede the observation of the QSH phase in future experiments. 

\begin{figure*}
   \centering
   \includegraphics[width=17cm]{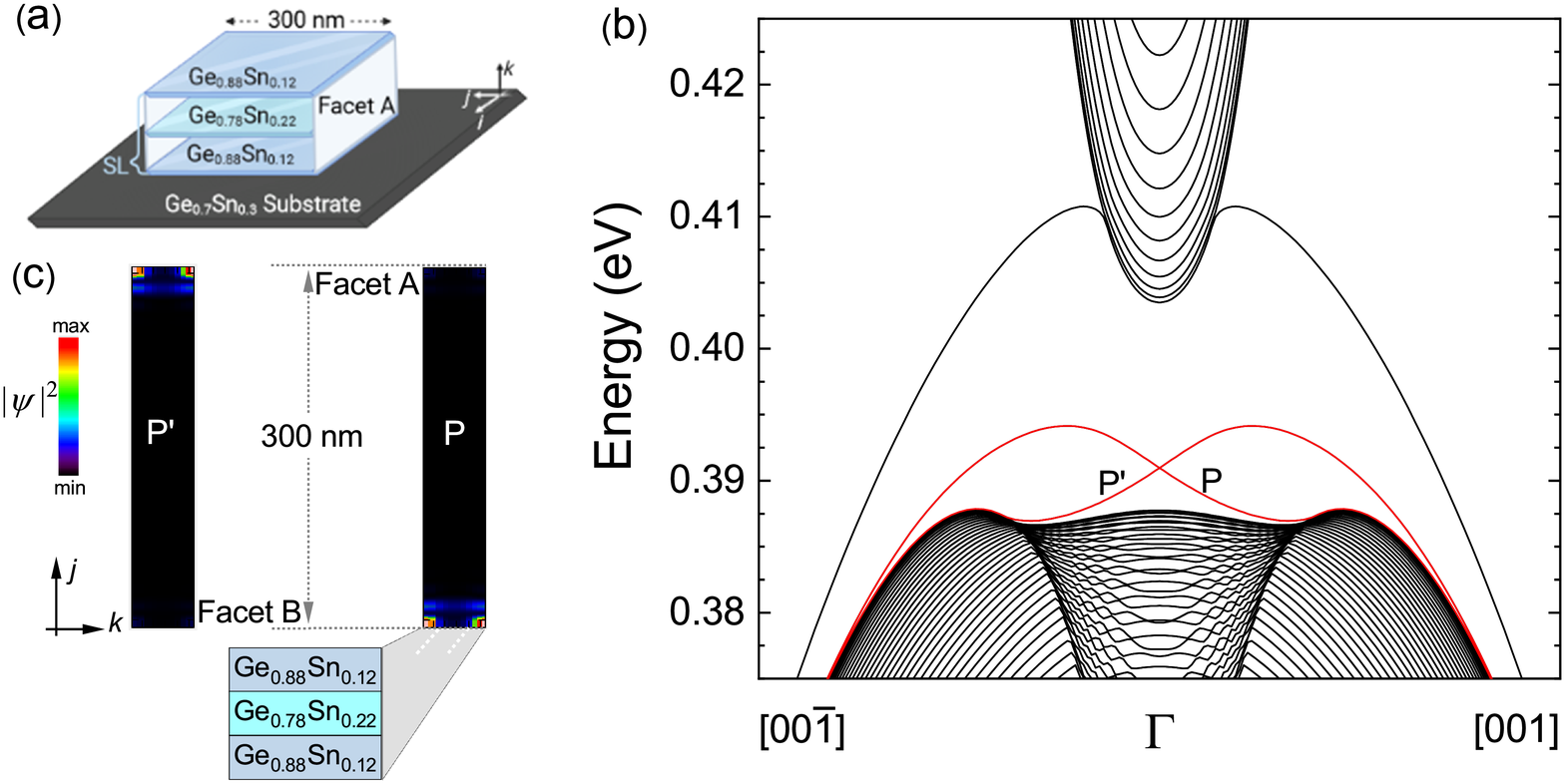}
   \caption{(a) Schematics of a $\mathrm{Ge_{0.78}Sn_{0.22}}$(25 nm)/$\mathrm{Ge_{0.88}Sn_{0.12}}$(25 nm) SL, which is fully strained with respect to a $\mathrm{Ge_{0.7}Sn_{0.3}}$ buffered substrate. The finite-size Hall bar has a width of 300 nm and extends along the \textit{j} direction (not to scale). (b) Band structure of the two-dimensional ribbon. The red lines highlight the topological edge states. (c) Colour-coded maps showing the probability density $\left|\Psi^2\right|$ associated to the surface states (P and $\mathrm{P^\prime}$).} \label{fig3}
\end{figure*}

For consistency with the literature, we consider a symmetric $\mathrm{Ge_{1-x}Sn_x/Ge_{1-y}Sn_y/Ge_{1-x}Sn_x}$ heterostructure. Without loss of generality, the two external identical $\mathrm{Ge_{1-x}Sn_x}$ films will be dubbed barriers, whereas the intermediate $\mathrm{Ge_{1-y}Sn_y}$ layer will be termed well. The line-ups at the heterointerface have been obtained by varying the strain in the epitaxial stack through the modification of the Sn molar fraction of the $\mathrm{Ge_{1-s}Sn_s}$ buffer. Specifically, we utilized elemental Ge and a $\mathrm{Ge_{0.70}Sn_{0.30}}$ alloy. The latter has been chosen as un upper limit to guarantee experimental feasibility using state-of-the-art deposition techniques. For the same reason, we restricted the barriers to a Sn-diluted regime, namely $x<12$\%. Even though in this work we focus on the experimentally accessible region of the parameter space, the $\mathrm{Ge_{1-y}Sn_y}$ well was allowed to freely sample the whole composition range ($0<y<1$) to give a better overview of all the domains potentially characterized by the sought-after BG. 
Fig.~\ref{fig1} summarizes the result that were obtained when opposite ends of the y range are considered \cite{supp}. As a guideline, Figure~\ref{fig1}a and ~\ref{fig1}b also report a schematic of the offsets at the conduction band (black lines) and at the top of the valence band (red lines). The strain dependence is shown in Fig. S3 \cite{supp}.

It is interesting to notice the rich atlas of line-ups unleashed in the $\mathrm{Ge_{1-x}Sn_x}$ system. Insights can be obtained by the gross features that exquisitely appear in Fig.~\ref{fig1}a-b. 
When the Sn molar fraction of the substrate is higher than $\sim 0.26$, a Sn bulk-like band structure can be achieved within the films (INV, grey in Fig.~\ref{fig1}a and b). However, when $s<0.26$, an inverted regime can be prominently observed when the Sn content of the $\mathrm{Ge_{1-y}Sn_y}$ wells exceeds $\sim 50$\%. Specifically, a region exhibiting an inverted type-I feature, as in HgTe/CdTe, opens below 60\%, while a distinct InAs/GaSb-like BG develops well above 60\% (Fig.~\ref{fig1}a). At a first glance, $\mathrm{Ge_{1-y}Sn_y}$ wells with a Sn molar fraction larger than 50\% can be regarded as strong candidates for hosting topologically nontrivial phases. It should be noticed, however, that although very promising, alloys of high Sn content are hard to synthesise. In the low Sn content scenario, namely when $\mathrm{Ge_{1-y}Sn_y}$ wells with $y<0.5$ are considered, the strain imprinted by the substrate appears to restore the normal band sorting, featuring either a type-I (yellow area) or a staggered type-II (green area) alignment (Fig.~\ref{fig1}a). In the latter case, charges are set apart at opposite sides of the heterointerface, where a well-defined insulating gap is retained. Surprisingly, in this diluted alloy regime there exists a region in which our calculations predict the emergence of a unique BG. This occurs around $0.23<s<0.3$ and $0<y<0.4$ that is marked in red in Fig.~\ref{fig1}a-b. Such a band has the advantage of having a direct gap in both the well and substrate and of being manifested in heterostructures that are easier to grow through epitaxial techniques. 
Another central finding regarding the interface alignments is the energy difference $\Delta E$ calculated between the bottommost conduction band edges and the topmost valence band, which is established across the junction. Fig.~\ref{fig1}c-d demonstrate negative (positive) $\Delta E$ when the inverted (normal) band ordering sets in. Interestingly, $\Delta E$ can be extraordinarily large as these gaps possess remarkable amplitudes in the range of several hundreds of meV. This offers indeed plausible prospects for topological applications well above cryogenic temperatures.

\begin{figure*}
   \centering
   \includegraphics[width=17cm]{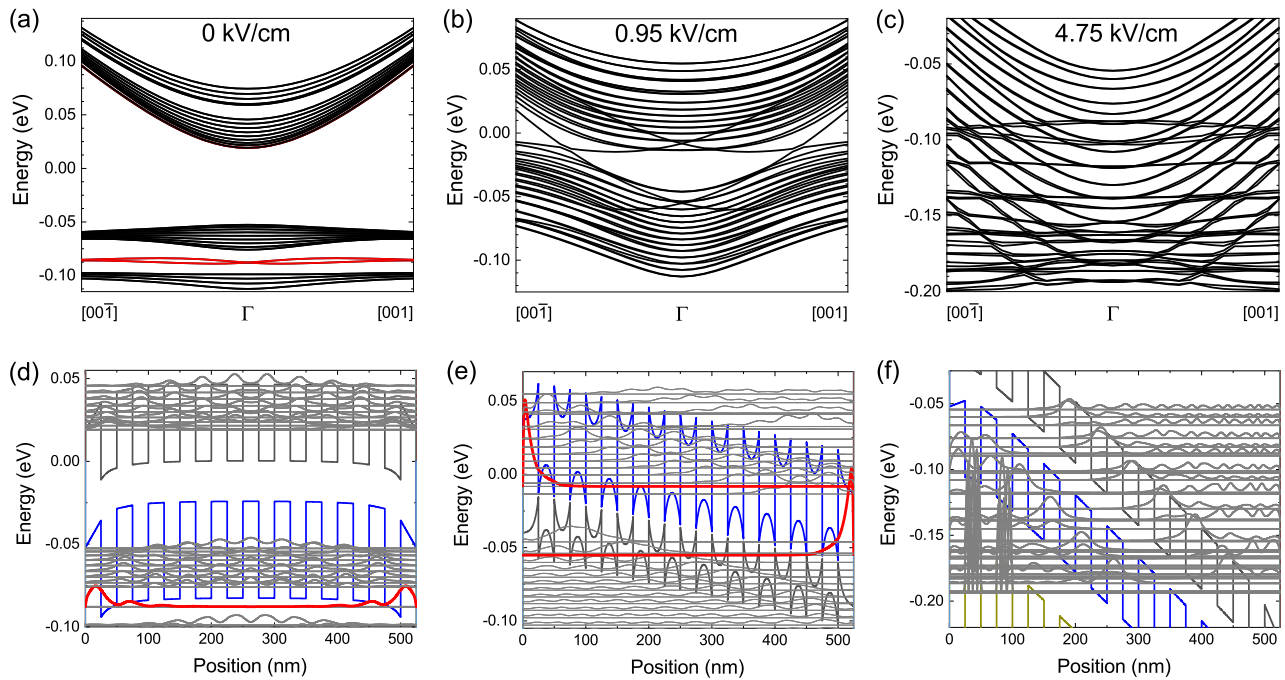}
   \caption{Energy spectrum of a ten-fold $\mathrm{Ge_{0.88}Sn_{0.12}}$(25 nm)/$\mathrm{Ge_{0.78}Sn_{0.22}}$(25nm) SL grown on a $\mathrm{Ge_{0.75}Sn_{0.25}}$ buffer. The SL is unbiased (a) or under an electric field of 0.95 (b) or 4.75 kV/cm (c). The red branches indicate the topological states (see also \cite{supp}). Band edges and squared wave functions for the lowest energy subband states at the centre of the Brillouin zone refers to an unbiased device (d) or when electric fields of 0.95 (e) or 4.75 kV/cm (f) are applied.} \label{fig4}
\end{figure*} 

In the following we turn our attention to a $\mathrm{Ge_{0.88}Sn_{0.12}/Ge_{0.78}Sn_{0.22}}$ multilayer stack and analyse the evolution of the electronic states as a function of the thickness of the layers as well as strain relative to the substrate. In Fig.~\ref{fig1} such heterointerface lays within the parameter space previously disclosed by the model-solid theory as the source of a BG. 

We now leverage a multiband $k \cdot p$ model \cite{birner07} to unveil the richness in terms of quantum phases and to fully capture the essential physics of this new and intriguing $\mathrm{Ge_{1-x}Sn_x}$ platform. At first, we investigate an infinite periodic $\mathrm{Ge_{0.88}Sn_{0.12}/Ge_{0.78}Sn_{0.22}}$ SL. We set the thickness of the $\mathrm{Ge_{0.78}Sn_{0.22}}$ well at 10 nm, while the width of the $\mathrm{Ge_{0.88}Sn_{0.12}}$ barrier is maintained relatively large, e.g., 25 nm, to ensure a reduced penetration of the wavefunctions throughout the whole SL period.
The energy spectra calculated by the exact diagonalization of the $k \cdot p$ model along the well plane when the virtual substrate terminates with $\mathrm{Ge_{0.75}Sn_{0.25}}$ (Fig.~\ref{fig2}a), and $\mathrm{Ge_{0.71}Sn_{0.29}}$ (Fig.~\ref{fig2}b) suggest strain to be a pivotal driving force for the closure of the band gap and the likely cause of the hybridization between electron and hole states. Fig.~\ref{fig2}c summarizes the calculated shift of the energy levels of the SL as a function of the Sn content of the buffer. A striking transition from a trivial to a topological insulator can be clearly seen to occur at the critical value $s_c = 0.27$. The strain introduced by the substrate provokes indeed an opposite behaviour for the electron and hole subbands, yielding a crossover that drives the SL structure into an inverted scheme when $s>s_c$ (Fig.~\ref{fig2}c). Notice that in the topological regime, that is, above $s_c$, the opening of the bulk insulating gap can be tuned in a very drastic way, as it remarkably approaches a magnitude of more than 1 eV in the limiting, although unpractical, case of growth on almost pure Sn buffers. We can find reassurance on these results by the agreement with the $\Delta E$ band offsets disclosed by the model-solid theory. Remarkably, the amplitude of the gapped bulk of our $\mathrm{Ge_{1-x}Sn_x}$ proposal is exceedingly large compared to conventional 2D QSH insulators, whose value typically lays below 10 meV \cite{bernevig06,liu08, miao12, zhang13}, and compares well with atomically thin materials \cite{lodge21}.

Next, we consider a $\mathrm{Ge_{0.7}Sn_{0.3}}$ buffer whose Sn content is above $s_c$. Fig.~\ref{fig2}d shows no crossing points in the investigated range of the well thickness, thereby indicating that width fluctuations cannot destroy the attained normal and nontrivial insulator regimes. 
In the following, we focus on the study of the edge state spectrum to explicitly demonstrate the emergence of the gapless helical states. To achieve this purpose, we set an equal thickness of 25 nm for both the well and barrier layers and break the translational symmetry of the SL by opening free surfaces along the crystallographic direction \textit{j} lying in the QW plane, while keeping periodic boundary conditions along \textit{k} as shown in Fig.~\ref{fig3}a. The lateral dimension of the nanoribbon along \textit{j} is set to 300 nm to minimize interactions between states laying at opposite facets of the Hall bar (Fig.~\ref{fig3}a). 
In figure~\ref{fig3}b we report the electronic band structure in the neighbour of the $\Gamma$ point along the \textit{i} direction chosen parallel to the $[001]$ crystallographic axis. Fig.~\ref{fig3}b shows two linearly dispersing energy states (red) appearing inside of the gapped bulk (black) and crossing each other at the $\Gamma$ point. Fig.~\ref{fig3}c reports the square modulus of the wavefunction associated to these interesting states, which are confined along the growth axis \textit{k} and behave as plane wave in the \textit{i} direction, which is in the SL plane. Above all, their vanishing square amplitude in the bulk of the SL, that is, along \textit{j}, openly unveils that these are Kramers$^\prime$s pair tightly localized at the opposite sides of the nanoribbon. This eventually provides a compelling proof that the QSH phase indeed exists in group IV heterostructures. Calculations suggest that degenerate bulk states can possibly occur in these heterostructures, thus indicating that, as for bulk HgTe and $\alpha$-Sn, special care will be needed to single out chiral edge states in future experimental works \cite{khaetskii22}.

The unique coexistence and concomitant spatial separation of 2D electron and hole gases in the $\mathrm{Ge_{0.88}Sn_{0.12}/Ge_{0.78}Sn_{0.22}}$ QSH insulator spontaneously establishes a dipole layer across the interface. This raises the attractive prospect of controlling the hybridization through electric fields. We discuss this possibility by focusing on the model $\mathrm{Ge_{0.88}Sn_{0.12}}$(25 nm)/$\mathrm{Ge_{0.78}Sn_{0.22}}$(25nm) SL in the form of a finite size stack comprising 10 periods, which we have studied for different Sn concentrations covering the range were a phase transition can occur. This allows us to identify, via self-consistent $k \cdot p$ calculations, the impact on the QSH phase of an external electrical bias, as we applied ohmic contacts at the two closing ends of the SL right along the \textit{k} growth axis.

As we scan the Sn molar fraction of the substrate at zero bias, we recover the opening of a minigap within the valence band with the distinct formation of two complementary localized conductive edge states that become more pronounced as the Sn content, i.e., strain, increases (see also Fig S4 and Ref. \cite{belopolski17}). The resulting density distribution further supports our conclusions as we find these states to be confined at the top and bottom surfaces when $s>0.2$. It is worth noting that in this calculation the topological states emerge solely because of the finite dimension of the SL along the growth direction. That is distinct from the results obtained for the previously discussed infinite SL in the nanoribbon geometry (Fig.~\ref{fig3}), thereby reinforcing the case for $\mathrm{Ge_{1-x}Sn_x}$-based heterostructures as robust hosts of the nontrivial QSH phase.
We therefore study the influence of an external bias on the topological structure focussing on a SL deposited on a $\mathrm{Ge_{0.75}Sn_{0.25}}$ buffer layer. The results are summarized in Fig.~\ref{fig4}. Charge confinement manifests itself at 0 bias as a bending in a neighbourhood of the final layers of the SL, while the profile at the centre remains flat.

By applying a gentle positive bias, the edge states are more tightly confined and even more peaked at the free surfaces. At the same time, the external electric field drives the accumulation of charge density at each well/barrier interface as demonstrated by the characteristic bending profile, which is seemingly different from the flat band condition. Notably, the further increase of the polarization causes the edge states to fade away (see panels c and f), inducing a phase transition of the SL from a QSH to a trivial insulator. A finding that provides a novel route towards the future implementation of topological field effect devices.

In conclusion, we unveiled quantum phases in silicon-compatible materials beyond conventional systems. Our platform naturally permits strain and electric field control of the quantum phases on Si. This concept is highly flexible and can have a longstanding impact, stimulating new research directions towards Si topology. 

\begin{acknowledgments}
We acknowledge C. Colombo and A. Filippi for technical assistance with model-solid theory and $k \cdot p$ calculations, respectively. F.M.-A. would like to acknowledge Science Foundation Ireland grant SFI/19/FFP/6953.
\end{acknowledgments}


%


\end{document}